\documentclass[reprint,aps]{revtex4-1}
\usepackage{graphicx}
\usepackage{bm}
\usepackage{color}

\newcommand{\GeV}{\makebox{ GeV}}
\newcommand{\beq}{\begin{equation}}
\newcommand{\enq}{\end{equation}}
\newcommand{\beqa}{\begin{eqnarray}}
\newcommand{\beqast}{\begin{eqnarray*}}
\newcommand{\enqa}{\end{eqnarray}}
\newcommand{\enqast}{\end{eqnarray*}}
\newcommand{\nn}{\nonumber}

\newcommand{\req}[1]{(\ref{#1})}

\newcommand{\cT}{{\cal T}}

\newcommand{\al}{\alpha}
\newcommand{\be}{\beta}
\newcommand{\ga}{\gamma}
\newcommand{\de}{\delta}
\newcommand{\ep}{\epsilon}
\newcommand{\ze}{\zeta}

\newcommand{\la}{\lambda}

\newcommand{\si}{\sigma}

\newcommand{\ph}{\phi}

\newcommand{\om}{\omega}

\def\GeV{\nobreak\,\mbox{GeV}}

\begin{document}


\bigskip

\title{   Scale Dependent Pomeron Intercept in Electromagnetic Diffractive Processes  }
\author{H. G. Dosch}
\affiliation{Institut f\"ur Theoretische Physik, Universit\"at Heidelberg \\
Philosophenweg 16, D-69120 Heidelberg, Germany     }
\author{E. Ferreira}
\affiliation{Instituto de F\'{\i}sica, Universidade Federal do Rio de Janeiro \\
C.P. 68528, Rio de Janeiro 21945-970, RJ, Brazil }



\begin{abstract}

We test the hypothesis that diffractive scattering in the
perturbative and non-perturbative domain is determined by the
exchange of a single pomeron with a scale dependent trajectory.
Present data on diffractive vector meson production are well
compatible with this model and recent results for $J/\psi$
photoproduction at LHC strongly support it. The model is inspired
by concepts of gauge/string duality applied to the pomeron.

\end{abstract}

\maketitle

\section{Introduction\label{intro}}
 Diffractive  processes involving virtual photons show a remarkable feature: the higher the photon virtuality $Q^2$,
 the faster is the increase of the cross sections with energy.
This feature is well understood in perturbative QCD, where the
evolution equations in $Q^2$
\cite{Gribov:1972ri,Altarelli:1977zs,Dokshitzer:1977sg} predict
such  behaviour. Microscopically, the rising increase in energy
can be traced back to the increase of the gluon density with
higher resolution. This
 specific feature of the energy dependence, however, is less  easily explained  in Regge theory.
 In purely  hadronic diffractive processes the energy dependence of the scattering amplitude is determined by the pomeron trajectory to be $W^{2 \al_P(t)}$ where $W$ is the cm energy,
$\al_P(t) = \al_P(0)  + \al'_P  \,t $ is the pomeron trajectory,
  and $t$ is the squared momentum transfer. Based on a large amount  of hadronic diffractive  data, Donnachie and Landshoff \cite{Donnachie:1992ny} proposed a general description with a
hypercritical pomeron intercept $\al_P(0) \approx 1.09$ and a
slope  $\al'_P = 0.25 \GeV ^{-2}$.

\begin{figure}
\begin{center}
 \includegraphics*[width=7cm]{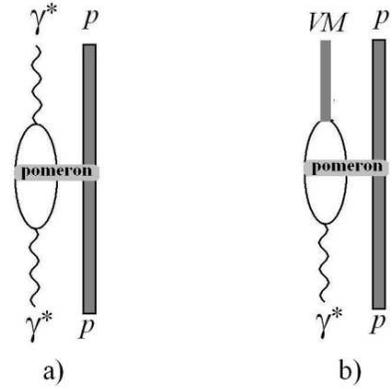}
\end{center}
 \caption{ \label{dipole} Electromagnetic diffractive processes in the dipole model governed by pomeron exchange:   \quad a)~$\ga^*$ scattering;
b)~electroproduction of vector mesons (VM) . }
\end{figure}

  Electroproduction processes can be related to purely hadronic interactions
through the assumption that the photon-hadron interaction occurs
via the interaction of the target hadron with a quark-antiquark
pair, as illustrated in  Fig.\ref{dipole}. According to  this
model,  diffractive electromagnetic processes are described as
purely hadronic processes, with  energy dependence  governed by
the pomeron trajectory. This approach, commonly called dipole
picture~ \cite{Nikolaev:1991et}, has been tested in many analyses
and applications. Although there are certain limitations to this
approach \cite{Ewerz:2006vd,Ewerz:2007md} it is intuitive and
phenomenologically very successful.

 On the other hand,   by summing up leading log terms in perturbative  QCD,  a larger pomeron  intercept was found (BFKL-Pomeron)~\cite{Fadin:1975cb,Kuraev:1977fs,Balitsky:1978ic,Fadin:1998py,Ciafaloni:1998gs}.
 Donnachie and Landshoff \cite{Donnachie:1998gm} extended the pomeron concept and assumed that electromagnetic diffractive  processes are determined by two pomerons, a soft (hypercritical) one with an intercept of about 1.09 and a hard one with an  intercept of about 1.4. This idea has been applied in many electroproduction processes, where the residues of the pomerons were essentially determined  by the size of the scattered objects and hence in a given model the energy dependence was universally fixed by a superposition of the two pomeron contributions. In this way a comprehensive description of  proton structure functions, vector meson production
and $\gamma^*-\gamma^*$ scattering could be achieved  in the full
energy  range accessible at HERA
\cite{Dosch:1994ym,Dosch:1997nw,Kulzinger:1998hw,Donnachie:1999kp,
Donnachie:2000px,Dosch:2002ig,Dosch:2006kz,Baltar:2009vp}.

Recent experiments on $J/\psi$ photoproduction at LHC
energies~\cite{LHCb(14),Alice(14)}, however,  show  that a single
power, corresponding to a  pomeron intercept at about 1.17,
describes very well the energy dependence in the range from 20 GeV
to 1 TeV. This result  strongly supports the concept of  a single
pomeron exchange.

The AdS/CFT  correspondence
\cite{Maldacena:1997re,Gubser:1998bc,Witten:1998qj} has opened
interesting new aspects to pomeron physics
\cite{Brower:2006ea,Hatta:2007he}. In a very simple model  in  an
ultralocal approximation, discussed at the beginning of \cite{Brower:2006ea}   it was shown
that the pomeron trajectory depends on the holographic variable
$z$ \beq \label{alp} \al_P(t,z) = \al_P(0)  + \al'_P(z) \,t  ~,
\enq with \beq \label{depslope}
  \al'_P(z) = \al'\,\frac{ z^2 }{R^2}~,
\enq where $\al'$ is a constant relevant for scattering of the strings in the higher dimensional space  and $R$ is the $AdS$ scale. The holographic variable $z$ used here is related to the $r$ in \cite{Brower:2006ea} by $z= R^2/r$. In this simple model the
intercept value is $\al_P(0) = 2$~\cite{Brower:2006xx}.

With the assumption of a single
pomeron the value  of the intercept   should be
nearly one in order to describe purely hadronic diffraction processes, whereas experimentally it is distinctly larger for
diffractive processes with high photon virtuality.  Therefore, in
a  bottom-up approach to the problem, we are led to extend the
concept of a scale dependent pomeron slope, Eq.\req{depslope},  to
that of  a scale dependent intercept, $\al_P(0,z)$.  In order to
test this  assumption quantitatively, we use in the present work a
phenomenologically very successful   approach to AdS/CFT duality,
namely light front holographic QCD, developed by  de T\'eramond
and Brodsky \cite{deTeramond:2008ht} (for a review see
\cite{Brodsky:2014yha}). One essential point is the discovery that
the AdS bound state equations are identical with the light front
(LF) bound state equations, if one identifies the holographic
variable $z$ with the boost invariant light-front separation
$\zeta$ between the quark and the antiquark inside a hadron
(details are given in the next section). It has been  shown that
the bound state wave functions  obtained in  light-front
holographic QCD give  a very good fit to rho-meson
electroproduction \cite{Forshaw:2013oaa}, and  it was shown
\cite{Dosch:2006kz} that the light-front wave functions \cite{lep80}
 are indeed very appropriate for the description of electroproduction not only  of $\rho$ mesons, but also of other vector mesons as $J/\psi$ and $\Upsilon$. In the present work we
use the coincidence of the holographic variable $z$ with the boost
invariant light front separation $\zeta$ in order to determine the
scale, on which the  pomeron intercept is assumed to depend, for
different diffractive processes.
 We proceed in the following way. First  the scale for different processes is determined and then  the dependence of the pomeron intercept is extracted from an analysis of the energy dependence of the structure functions~\cite{H1(10),Radescu:2013mka} on the scale. This intercept  determines
the energy behaviour of all processes with the same scale, and
the energy dependence of different processes, as $ \ga^*\, p$
scattering  and vector meson electroproduction, can be related.

 Our paper is organized as follows: In Sect.\ref{scale_sec} we determine the scale for the different processes and give interpolation formul\ae \quad  which relate the scale to the photon virtuality; in Sect.\ref{pomeron_sec} the pomeron intercept as function of the scale and the photon virtuality is derived  from the energy dependence of the structure functions. In Sect.\ref{data}  we present comparison with experiments and finally we summarize and discuss our results in Sect.\ref{final}.

\section{Fixing the scale for different processes  \label{scale_sec}}

Vector meson production and $\gamma^* p$ scattering are the best
investigated electromagnetic diffractive processes. In the dipole
model, for  a fixed  cm energy $W_0=\sqrt{s_0}$,of the $\gamma^*$ and the target
the forward scattering amplitude is generically given by~\cite{Nikolaev:1991et} \beq
\label{amp} \cT_0= i \int_0^\infty d \zeta \, \int_0^1  du
\frac{\zeta }{u \bar u} ~ \sigma(u,\zeta) ~ \rho(Q^2,u,\zeta) ~ ,
\enq where $u$ and $\bar u =1-u$ are the longitudinal  momentum
fractions of the dipole  constituents, and
\begin{equation}
 \zeta= \sqrt{u \, \bar u\;} \, r_\perp   ~ ,
\end{equation}
$r_\perp$ being their  transverse separation.

 The quantity
$\sigma(u,\zeta)$ is the dipole cross section and
 $\rho(Q^2,u,\zeta)$
 the overlap  of the wave functions
of the virtual photon and  of the diffractively produced particle. Eq. \ref{amp} determines the Regge residue and  the energy dependence of the forward amplitude in the Regge model
is  $\cT=\cT_0\, (s/s_0)^{\al_{\rm P}  }$.

The overlap, written generically as \beq \rho_{\ga^*,\it fs}
(Q^2,u,\zeta)= \psi^*_{\it fs}(u,\zeta) \;
\psi_{\gamma^*}(Q^2,u,\zeta), \enq
 is represented  diagrammatically in  Fig.\ref{dipole}. Here
 $\psi_{\gamma^*}(Q^2,u,\zeta)$ is the    hadronic wave function of the incident photon, and
$\psi_{\it fs}(u,\zeta) $  is the wave function of the final
state; for vector meson production it is the vector meson wave
function, while for    $\gamma^*$ scattering  it  is the outgoing
hadronic photon wave function.

For the overlap functions we present below the simple and
phenomenologically successful forms, which  at fixed cm energy W
describe  very satisfactorily the $Q^2$ dependence of the
different processes
 \cite{Kulzinger:1998hw,Donnachie:2000px,Dosch:2002ig,Baltar:2009vp}. Conservation of
$s$-channel helicity  is assumed  \cite{Gilman:1970vi}.

For $\ga^*\, p$ scattering with transverse polarization we write
  \begin{eqnarray}
&& \rho_{\ga^*\ga^*\pm1}(Q^2,u,\zeta) = \hat e_f^2\frac{6 \al}{4 \pi^2} \\
&&  \left[ (Q^2 \,u(1-u) + m_f^2) (u^2 +(1-u)^2 )\, K^2_1(\hat \ep
\zeta)+ m_f^2\,K^2_0(\hat \ep \zeta)\right] \nonumber
\end{eqnarray}
and for longitudinal polarization
\begin{eqnarray}
&& \rho_{\ga^*\ga^*0}(Q^2,u,\zeta)  \\ \nonumber && =\hat
e_f^2\frac{12 \al}{4 \pi^2} ~  Q^2 \,u^2(1-u)^2 \, K^2_0(\hat \ep
\zeta)  ~ ,
\end{eqnarray}
where
\begin{equation}
 \hat \ep = \sqrt{Q^2+ \frac{m_f^2}{u(1-u)} }  \ ~ ,
\end{equation}
and with $m_f$  representing the mass of the quarks forming the
dipole in the LHS of  Fig. \ref{dipole}.

The overlap functions connecting  $\gamma^*$ and   vector mesons
are for transverse polarization
\begin{eqnarray}
&& \rho_{\ga^*,V;\pm1}(Q^2,u,\zeta) = \hat e_V \frac{\sqrt{6 \al}}{2 \pi}  \\
&& \left[ 4 \hat \ep\, \zeta  \, \om^2 (u^2 +(1-u)^2 )\, K_1(\hat
\ep \zeta) + m_f^2\,K_0(\hat \ep \zeta) \right]
\ph_{\omega}(u,\zeta) ~ ,   \nonumber
\end{eqnarray}
 and for longitudinal polarization
\begin{eqnarray}
&& \rho_{\ga^*,V;0}(Q^2,u,\zeta) \\
&& = 16 \hat e_V \frac{\sqrt{3 \al}}{2 \pi}\,  \om \, Q \, u^2
(1-u)^2\,K_0(\hat \ep \zeta) ~  \ph_{\omega}(u,\zeta) ~ ,
\nonumber
\end{eqnarray}
 with   mass $m_f$ for  the quarks constituting the vector meson. $\ph_{\omega}(u,\zeta)$ is
the Brodsky-Lepage (BL) ~\cite{lep80} wave function  parameter
$\omega$, written \beq \label{BL} \ph_{\omega}(u,\zeta) =
\frac{N}{\sqrt{4 \pi}}\exp\left[-\frac{m_f^2(u-1/2)^2}{2 u (1-u)
\om^2}\right] \exp[-2 \om^2 \zeta^2] ~. \enq (In the wave
functions derived from light front holographic QCD
\cite{deTeramond:2008ht} an additional factor $\sqrt{u(1-u)}$
appears; the influence of this factor, however,  on the value of the scale is
negligible). For convenience, the values of $\omega$ in the   BL
wave function \req{BL}, determined by the electronic decay widths
\cite{Dosch:2006kz,Baltar:2009vp}  are given    in Table
\ref{WFparam}.

\begin{figure}
\begin{center}
\includegraphics*[width=7cm]{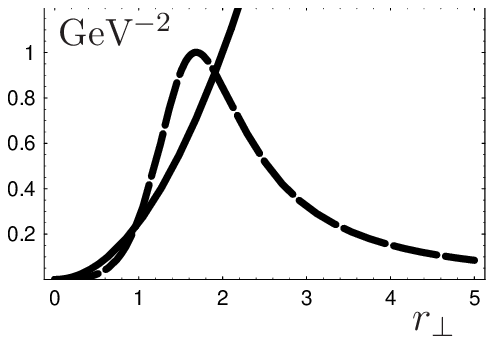}
\end{center}
\caption{ \label{dipole-cross}  Dipole cross sections. The solid
line represents the usual quadratic form \req{dip-cross}, and the
dashed line plots the form given in \cite{Jeong:2014mla}.}
\end{figure}

 Our final results do not depend strongly on the form of the dipole cross section, since it enters into the calculation for photon scattering and vector-meson production in the same way. The simplest choice is the   quadratic form
\beq \label{dip-cross} \sigma(u,\zeta) = c\, r_\perp^2 = c\,
\frac{\zeta^2}{u \bar u} ~ . \enq We also use a form obtained
directly from the data by a deconvolution of experimental data
~\cite{Jeong:2014mla,Ewerz:2011ph}. This dipole cross section
starts to grow with the third power of $r_\perp$ and decreases for
$r_\perp > 3 $ GeV$^{-1}$, as shown in Fig.\ref{dipole-cross}. The
value of the scale does not depend on the magnitude of the dipole
cross section but only on its form. In the following we shall use the quadratic form, unless  stated explicitely. 

In order to extract a fixed representative scale $\bar \zeta$ for a
specific process and a given $Q^2$ value, we  determine  the
$\zeta$ value where the overlap  is maximal, namely find $\bar
\zeta_{\rm max}$ that gives the maximum of the integrand $Y$ in
Eq.\req{amp} \beq \label{integrand} Y(Q^2,\zeta)=\int_0^1 du ~\
\frac{\zeta}{u \bar u} \sigma(u,\zeta) ~  \rho(Q^2,u,\zeta) ~ .
\enq

\begin{figure}
\begin{center}
\includegraphics*[width=8cm]{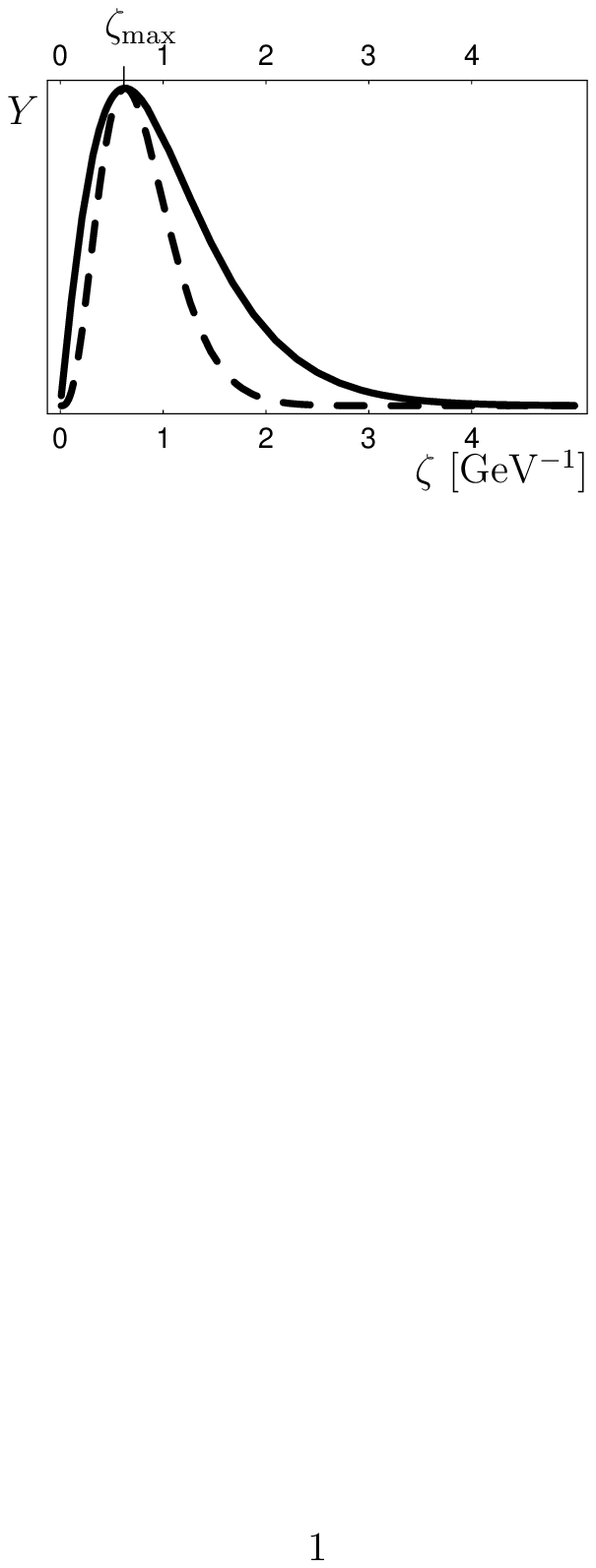}
\end{center}
\caption{ \label{overlaps} The function $Y$,  Eq.
(\ref{integrand}) , for $\rho_{\ga^* ,\ga^*,1}$  at $Q^2=1.5$
\GeV$^2$ (solid curve) and for   $\rho_{\ga^*\, \rho,1}$ at $Q^2=
26$ \GeV$^2$ (dashed curve) for the quadratic dipole cross section
\req{dip-cross} as function of $\zeta$. The displayed $Q^2$ values
are chosen to yield  the same $\bar \zeta_{\rm max}$ values, and
the curves are normalized to equal height at $\bar \zeta_{\rm
max}$.}
\end{figure}

Typical forms of the function  $ Y(Q^2,\zeta)$  for two chosen
processes  with the quadratic dipole cross section and selected
$Q^2$ values are shown in Fig.\ref{overlaps}. As we are interested
in showing that   both overlap functions   have marked maxima
positions, but different shapes,  the curves are presented
normalized to 1 at the peak, and two $Q^2$ values are chosen that
show peaks at the same value of $\bar \zeta_{\rm max}$.
 The values represented in the figure are  $Q^2=1.5$ GeV$^2$ for $\rho_{\ga^* ,\ga^*,1}$
and $Q^2= 26$ GeV$^2$ for   $\rho_{\ga^*\, \rho,1}$ .

In order to obtain some control of the uncertainties of this
choice of a representative scale, we also use another intuitive
criterion.   We define $\bar \zeta_{\rm med}$ as the median value
of $\zeta$, determined  by
\begin{eqnarray}  \label{median}
&& \int_0 ^{\bar \zeta_{\rm  med}} d\zeta \int_0^1 d u \,  \frac{\zeta}{u \bar u} \sigma(u,\zeta)  \rho(Q^2,u,\zeta)  \\
&& = \int_{\bar \zeta_{\rm med}}^\infty  d\zeta \int_0^1 du
\frac{\zeta}{u \bar u} \sigma(u,\zeta)  \rho(Q^2,u,\zeta) ~.
\nonumber
\end{eqnarray}

The cross sections of longitudinal and transverse photons are
added incoherently, and we therefore have fitted independently the
scales for both polarizations and for their weighted average, for
the different  processes and interesting $Q^2$ values.

 The overlap functions $\rho(Q^2,u,\zeta)$ depend on  the quark masses.
For diffractive production of heavy vector mesons we use the
$\overline{MS}$ masses \cite{Agashe:2014kda}: $m_c=1.28$ GeV and
$m_b=4.18$ GeV. For $Q^2 =0$  the overlap  diverges
logarithmically with vanishing quark mass and therefore special
constituent mass values have to be assumed. In order to reduce
model dependence,
 we have for light meson production determined the scale only for $Q^2 \geq 1$, where the dependence
on quark masses is weak and the current quark masses,
 $m_u \approx m_d \approx  0$, $m_s=0.1 $ GeV can safely be chosen.
For hadronic processes involving light quarks, the scale at
$Q^2=0$  is fixed by the confinement scale and therefore we have
there  the purely hadronic pomeron intercept $\al_{\bf P} \approx
1.09$.

\begin{figure}
\begin{center}
\includegraphics*[width=8cm]{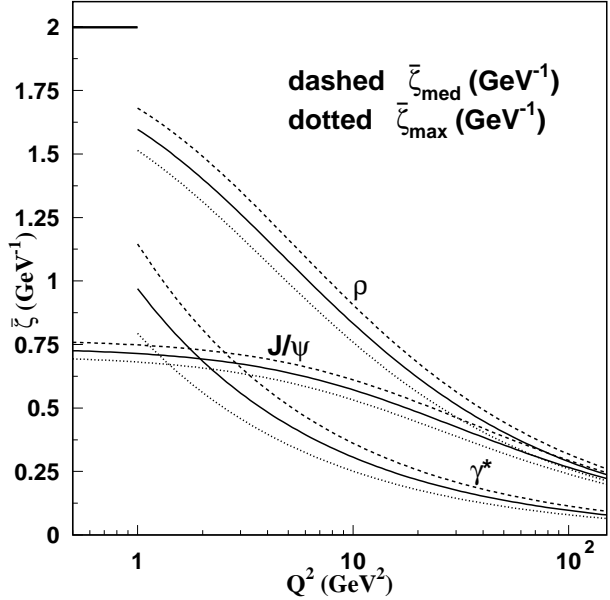}
\end{center}
 \caption{ \label{scales_fig} Plots of   functions
  $\bar \zeta_{\rm max} (Q^2)$ and $\bar \zeta_{\rm med }(Q^2)$
  that convert $Q^2$ into the scale variables that are
    used to  describe  different diffractive
   processes of  $\rho$ and $J/\psi$  vector meson production
and  photon scattering. The solid line represents their average
$\bar \zeta$ given in Eq.\ref{zetabar}.
 }
 \end{figure}

  Fig.\ref{scales_fig} shows the functions  $\bar \zeta_{\rm max} (Q^2)$ and
$\bar \zeta_{\rm med }(Q^2)$ and their average $\bar \zeta (Q^2)$
for  $\rho$ and $J/\psi$  vector meson production and for  photon
scattering.

\begin{table}[h]
\begin{center}
  \begin{tabular}{|l | c c | c c |}\hline
\multicolumn{1}{|l}{} &  \multicolumn {2}{|c|} { Transverse} &
          \multicolumn {2}{|c|} { Longitudinal}\\
&$~ \om$ & $N$ &$~ \om$ & $N$  \\
&   (GeV) & $ $    & (GeV)& $  $ \\  \hline
$\rho(770)$ &$0.2809 $  &$2.0820$&$0.3500$   &$1.8366$ \\
$\omega(782)$   &$0.2618 $  &$2.0470$&$0.3088$   &$1.8605$\\
$\phi(1020)$    &$0.3119 $  &$1.9201$&$0.3654$   &$1.9191$ \\
$J/\psi(1S)$    &$0.6452 $  &$1.4752$&$0.7140$   &$2.2769$ \\
$\Upsilon(1S)$  &$1.3333 $  &$1.1816$&$1.3851$   &$2.7694$ \\
\hline  \end{tabular}  \end{center} \vspace{-5mm}
  \caption{ \label{WFparam}  Parameters of the Brodsky-Lepage (BL) vector meson wave
functions (\ref{BL}), taken from
\cite{Dosch:2006kz,Baltar:2009vp}. }
\end{table}

 \section{Scale dependent pomeron intercept   \label{pomeron_sec} }

The best investigated electromagnetic diffractive process is
$\ga^*\,p$ scattering. It is usually presented as the DIS proton
structure function  $F_2(x,Q^2)$, which is related to the $\ga^*
\, p$ total cross section by
  \begin{equation}
F_2(x,Q^2) = \frac{Q^2}{4 \pi^2 \, \al} \si^{\ga^*p}  ~,
  \end{equation}
  with
\begin{equation}
   x=\frac{Q^2}{W^2 + Q^2 -m_p^2} ~ .
  \end{equation}
  Due to the optical theorem  the  total $\ga^*\, p$ cross section is proportional
to  the $\ga^*\, p$ forward scattering amplitude and its energy
behaviour is given by the pomeron intercept at $t=0$, with the
form $W^{2 [\al_{\rm P}(0)-1]}$.

  The structure function has  been fitted~\cite{Adloff:2001rw} with the power behaviour
\beq F_2(x,Q^2)= c ~ Q^2  \, x^{-\la(Q^2)} ~ , \enq with $\la(Q^2)
= 0.0481 \log\left[{Q^2}/{0.0853} \right]$ for $Q^2 \geq 3.5$
GeV$^2$, related to the pomeron intercept by ${\la =\al_{\bf P}
(0) -1}$ . We use the modification
\begin{equation}\label{laga} \al_{\rm \bf P}(0)-1= 0.0481 \log\left[\frac{Q^2+0.554}{0.0853}\right] ~ ,
\end{equation}
that is adjusted to give the intercept 1.09 at hadronic scales,
that is at $Q^2=0$.

    In order to relate the pomeron intercept with the scale $\bar \zeta$ one
inverts the scale function $\bar \zeta_\ga(Q^2)$, obtained  for
$\gamma^*\, p$ scattering according to the methods explained in
the preceding section; this yields the inverse function
$\overline{Q^2}_{\ga}\big(\bar \zeta\big)$.  One can then  calculate
the value of the intercept of vector meson production as function
of $Q^2$ from \req{laga} by inserting for $Q^2$ the value $\overline{Q^2}(\bar \zeta_{VM})$, where $\bar \zeta_{VM}$ is the  value
$\bar \zeta_{\rm VM}(Q^2)$ obtained for the vector meson at photon
virtuality $Q^2$.  We thus obtain  for the production of the
vector meson  VM the relation
\begin{equation}\label{laga2}
  \al_{\rm \bf P}(0)-1= 0.0481 \log\left[\frac{\overline{Q^2}_\ga\big(\bar \zeta_{\rm VM}(Q^2)\big)+0.554}{0.0853}\right] ~ .
\end{equation}

      It turns out that the intercept calculated for a specific process at  fixed $Q^2$
depends only weakly on  the method of its extraction, the
deviation of  $\al_P(0)-1$ obtained for both  procedures
$\bar\zeta_{\rm max}$ and $\bar\zeta_{\rm med}$
  deviates at most $\pm 5 \%$ from the mean value $\bar\zeta$.
 We therefore present in the following only the averaged results
\begin{equation} \label{zetabar}
\bar \zeta (Q^2)=\frac{1}{2}[\bar \zeta_{\rm max} (Q^2)+\bar
\zeta_{\rm med }(Q^2)]  ~ .
 \end{equation}
 The extreme choice of the dipole cross
section of \cite{Jeong:2014mla} leads  to an increase of
$\al_P(0)-1$ by less than 15 \%.

The results of the numerical analysis  show that for each process
$\ga^* \, p \to {\it fs} \, p$ with  final state fs    and given
polarization "pol", the average scale $\bar \ze (Q^2)$ can be very well
fitted by  a function of the simple form 
\beq \label{fit}
 \bar \zeta_{\it fs}(Q^2) = \frac{a_{\it fs, pol}}{\sqrt{Q^2 + b_{\it fs, pol}}} ~.
\enq 
The coefficients $a$ and $b$ for the different processes are
displayed  in Table \ref{res}. 

From this fit and Eq.\req{laga2} we
can obtain the pomeron intercept from $\ga^*$ scattering as a function of the scale
$\bar \zeta$.  We then have 
\begin{equation}\label{laga3}
   \al_{\rm \bf P}(0)-1= 0.0481 \log\left[\frac{a_{\gamma^*,\rm{pol}}^2/
{\bar \zeta}^2 -b_{\gamma^*,\rm{pol}}+0.554}{0.0853}\right] ~ ,
  \end{equation}
where $a_{\gamma^*,\rm{pol}}$  is the  coefficient in
Eq.\req{fit}  for $\ga^*\,p$ scattering, with the label pol
indicating transverse (T), longitudinal(L)  or total (tot) cross
sections (row  $\ga^* $ in Table \ref{res}). 

From this
equation we obtain the intercept for vector meson production as
function of $Q^2$ by expressing the scale $\bar\zeta$ through
Eq.\req{fit} for the specified meson VM
\begin{eqnarray}  \label{meq}
\delta(Q^2)&=& 4 ( \al_{\rm \bf P}(0)-1) = 0.472+ 0.1924 \times\\
&&\hspace{-1cm}  \log\left[\frac{a_{\gamma^*,\rm{pol}}^2}{a_{\rm
VM,pol}^2} (Q^2 -b_{\gamma^*,\rm{pol}}+b_{\rm VM,pol}) +0.554\right] ~.    \nonumber
 \end{eqnarray}

\begin{table}[h]
\begin{center}
\begin{tabular}{|c|ccc|ccc|}\hline
final state & \multicolumn{3}{|c|}{$a_{\it fs,\rm pol}$   }&\multicolumn{3}{c|}{$b_{\it fs,\rm pol}$ [GeV$^2$]}    \\
  {\it fs}  &trans&long&total&trans&long&total\\
\hline
$ \gamma^* $ &0.945&1.228&0.968 & -0.004&-0.003&-0.004\\
$\rho $&3.602&2.767&2.925&2.724&2.092&2.357\\
$ \phi$&3.651&2.800&3.022&3.308&2.581&2.351\\
$J/\psi$& 3.386&2.790&2.856&20.63&17.12&14.98\\
$\Upsilon$& 3.186&2.765&2.658&123.2&109.6&86.93\\
\hline
\end{tabular}
 \end{center}
\vspace{-5mm}
  \caption{ \label{res}  Coefficients of the numerical fits of  the average scale
$\bar\zeta_{\it fs} (Q^2) $, \req{zetabar}, for the processes $\ga^* { p} \to {\it fs} \, { p}$
 with Eq.\req{fit},  for use in longitudinal, transverse and total
(incoherent sum of the two cases)  cross sections. $fs=\gamma^*$ refers to 
$\gamma^* {\rm p}$   total cross section. The accuracy of
the fit is better than 1\% in the $Q^2$  range from 1 to 60
GeV$^2$ for photon scattering and $\rho,\,\phi$  production and
from  0 to  60 GeV$^2$ for $J/\psi$ and $\Upsilon$ production.
Remark: For $Q^2$ = 0 in $\rho$ and $\phi$  production,
the relevant scale is the hadronic scale,
 chosen as $\bar \zeta =2 \GeV^{-1}$, with a soft pomeron intercept
1.09.
 }
\end{table}

 The intercept at $t=0$ determines the energy behaviour of the forward scattering
amplitude (and therefore also of the total  $\ga^* \,p$ cross
section). For integrated elastic production cross sections one
also has to take into account the $t$ dependence of the
trajectory, which leads to a shrinkage of the diffraction peak.
For unpolarized elastic diffractive vector meson production, $\ga^*\,{\rm p }\to {\rm p}
\,{\rm [VM]}$, the differential elastic cross section in the Regge
model  is given by \beq
\frac{d\si}{dt}=\left(\frac{s}{s_0}\right)^{2 [\al_{\bf P}(t)-1]}
\be^2(t) ~ . \enq 
For fixed $W$ and $Q^2$ the $t$ dependence is
well approximated by an exponential, and we thus assume the
residue $\be(t) = \be_0 e^{B \,t/2}$ and
 $\al_{\bf P}(t) = \al_{\bf P}(0) + \al'_{\bf P}\, t $.  We then obtain for the
integrated   cross section
\begin{eqnarray}
&& \si_{\rm int} =\int_{-\infty} ^{0}  dt\,\frac{d\si}{dt} \\
&&  = \frac{\be_0^2}{B + 2\,\al'_{\bf P}\log(s/s_0)}\,
\left(\frac{s}{s_0}\right)^{2[\al_{\bf P}(0)-1]}\left(1 +
O(s^{-2})\right) ~.  \nonumber
\end{eqnarray}
The slopes observed in $d\sigma/dt$ in vector meson
electroproduction  \cite{Dosch:2006kz} are in the range of 5 to 10  GeV$^{-2}$.
With $2 \, \al'_{\bf P}/B \ll 1$,  the energy dependence of the
total cross section can be approximated by \beq \label{shrink2}
\si_{\rm int} \approx \frac{\be_0^2}{B}
\left(\frac{s}{s_0}\right)^{2[\al_{\bf P}(0) -  \al'_{\bf P}/B
-1]} ~ . \enq

Although  the present  data on $\al'_{\bf P}$ do not allow firm
conclusions~\cite{H1(10)}, it is certain that the effective powers $\delta_{\rm VM}$  that
fits experiments should be smaller  than the value $4 [\al_{\bf
P}(0) -1] $ obtained from the structure function. In the
simplified  model of Eq.\req{depslope}~\cite{Brower:2006ea}
discussed in the introduction, the slope of  the pomeron
trajectory decreases with decreasing scale 
\beq   \label{shrink1a}
\al'_{\rm P} =
\al' \frac{\bar \zeta^2}{\bar \zeta_{\rm conf}^2} ~ , \enq where
$\bar \zeta_{\rm conf}$ is the scale set by confinement, at which
$\al'_{\rm P}\approx 0.25 ~  $GeV$^{-2}$.
 Choosing   realistic values $\bar \zeta_{\rm conf}= 2 $  GeV$^{-1}$,
$B=5$ GeV$^{-2} $  we obtain a shrinking correction
\beq \label{shrink3} \frac{\al'_{\rm P}}{B} =
0.0125 \,\bar\zeta^2 = 0.0125 \, \frac{a_{\rm VM,pol}^2 }{Q^2 +b_{\rm
VM,pol}} ~, \enq 
and  for the power $\delta_{\rm int}$, applicable
to integrated elastic diffractive cross sections we have
\beqa
\label{delta-int}
\de_{\rm int}(Q^2) &=& \de - 4 {\al'}_{\rm P}/B = 0.472   \\
&&\hspace{-1.7cm} +0.1924
\log\Big[\frac{a_{\gamma^*,\rm{pol}}^2}{a_{\rm{ VM,pol}}^2 }\,
(Q^2 -b_{\gamma^*,\rm{pol}}+b_{\rm VM,pol})
+0.554\Big] \nn \\
&& \hspace{-1.7cm} -0.05 \; \frac{a_{\rm{VM,pol}}^2 }{Q^2 +b_{\rm{
VM,pol}}} ~ . \nn \enqa

\section{Description and prediction of diffractive data   \label{data} }

\begin{figure}
\begin{center}
\includegraphics*[bb= 17 150 525 654, width=8cm]{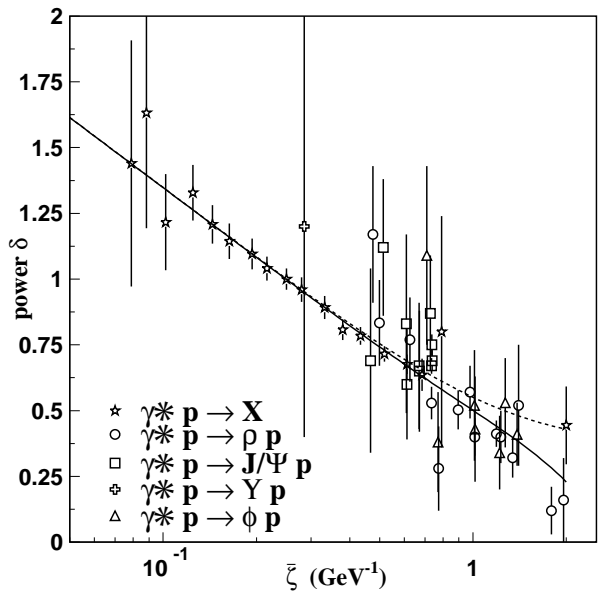}
\end{center}
\caption{ \label{all} Experimental values of $\delta = 4(\al_{\bf P}(0)-1) $
vs. the scale $\bar\zeta$ for different processes.  The
dashed line represents  the interpolation formula \req{laga}. The
solid curve takes into
account the effects of the shrinking in Eq.\req{shrink1a}.
 The stars are
obtained from measurements of the proton structure function and
total $\gamma^* $ p cross section \cite{Adloff:2001rw,Zeus(11)}.
References for the data on vector meson production are given in
detail in Fig.\ref{deltas}. }
\end{figure}

In Fig.\ref{all}   experimentally determined values of the power
$\delta=4\, \big(\al_{\rm P}(0)-1\big)$  for  different reactions are displayed   
against the  scale
$\bar\zeta$. The values for photon scattering are
deduced from measurements of the proton structure function and the 
total $\gamma^* $ p cross section \cite{Adloff:2001rw,Zeus(11)}. The 
experimental $\delta$ values for vector meson production are taken from:
a) $\rho$-production \cite{Zeus(98),Zeus(99),Zeus(07),H1(00),H1(10)};
 b) $\phi$-production \cite{Zeus(05),H1(10)};
c) $J/\psi$-production \cite{H1(06),H1(13),Zeus(02),Zeus(04)}; d)
$\Upsilon$-production \cite{Zeus(09)}.
They are given  for fixed $Q^2$, and the corresponding scale 
$\bar \zeta$ has been 
determined by Eq.\req{fit} with the constants from Table \ref{res}.  
The dashed line corresponds to the fit Eq.\req{laga} to the photon data  
with  $Q^2=0.968^2/{\bar \zeta}^2+0.004$. The solid line includes the shrinkage 
correction Eq.\req{shrink1a},\req{shrink3},\req{delta-int}
to be applied for the 
integrated cross sections of diffractive vector meson production. 

The errors
for vector meson production and correspondingly the fluctuations
are generally quite large, but the figure shows that the data  are
well compatible with  a common power behaviour, only dependent on
the $\bar\zeta$ scale, but not on the process.
 Future data in the TeV region with reduced errors may  provide 
decisive tests for the conjecture of a single pomeron with a scale dependent 
intercept governing the energy behaviour universally for all 
diffractive processes.

\begin{figure*}
\begin{center}
\includegraphics*[width=8cm]{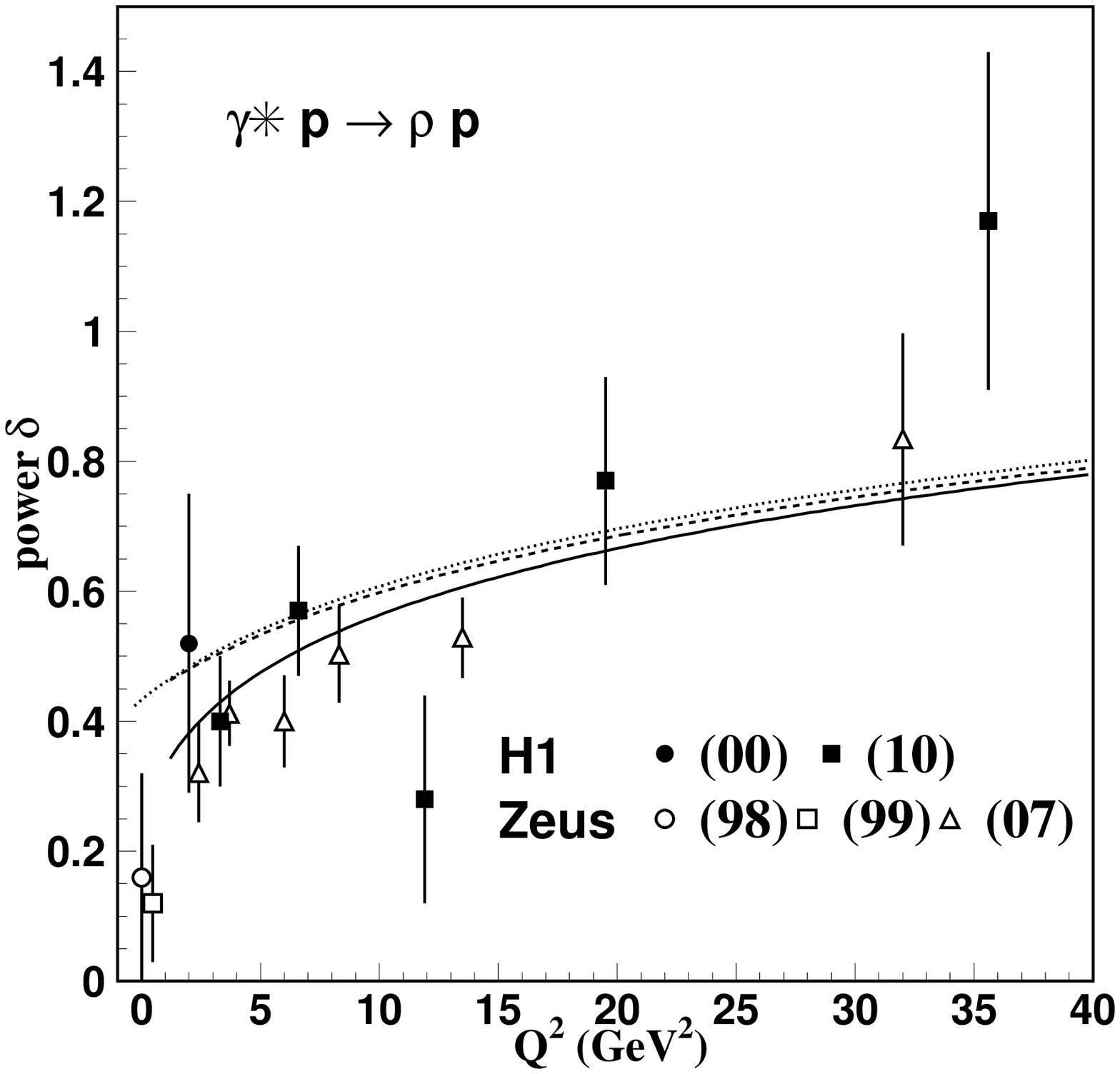}
  \includegraphics*[width=8cm]{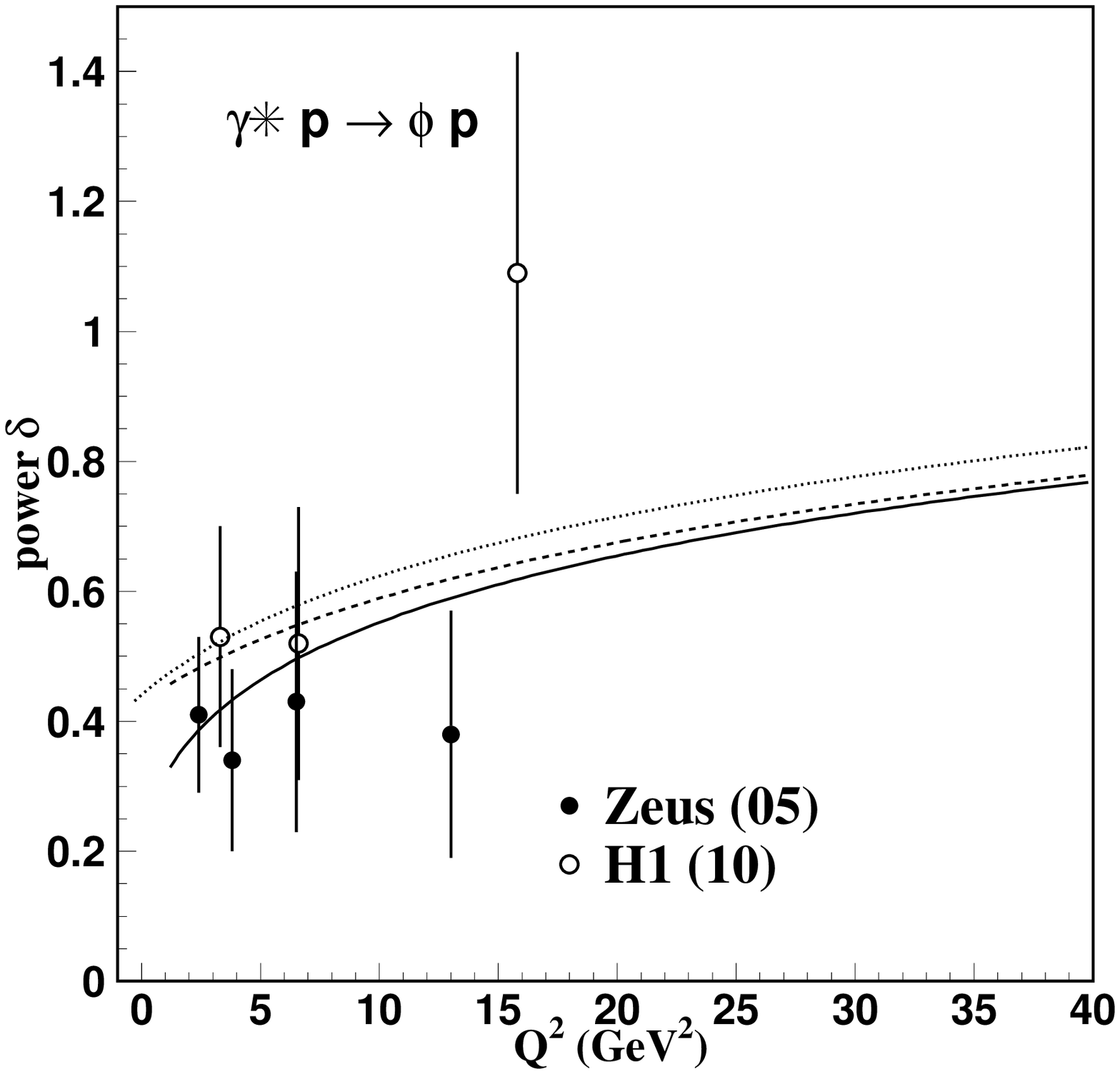}
  \includegraphics*[width=8cm]{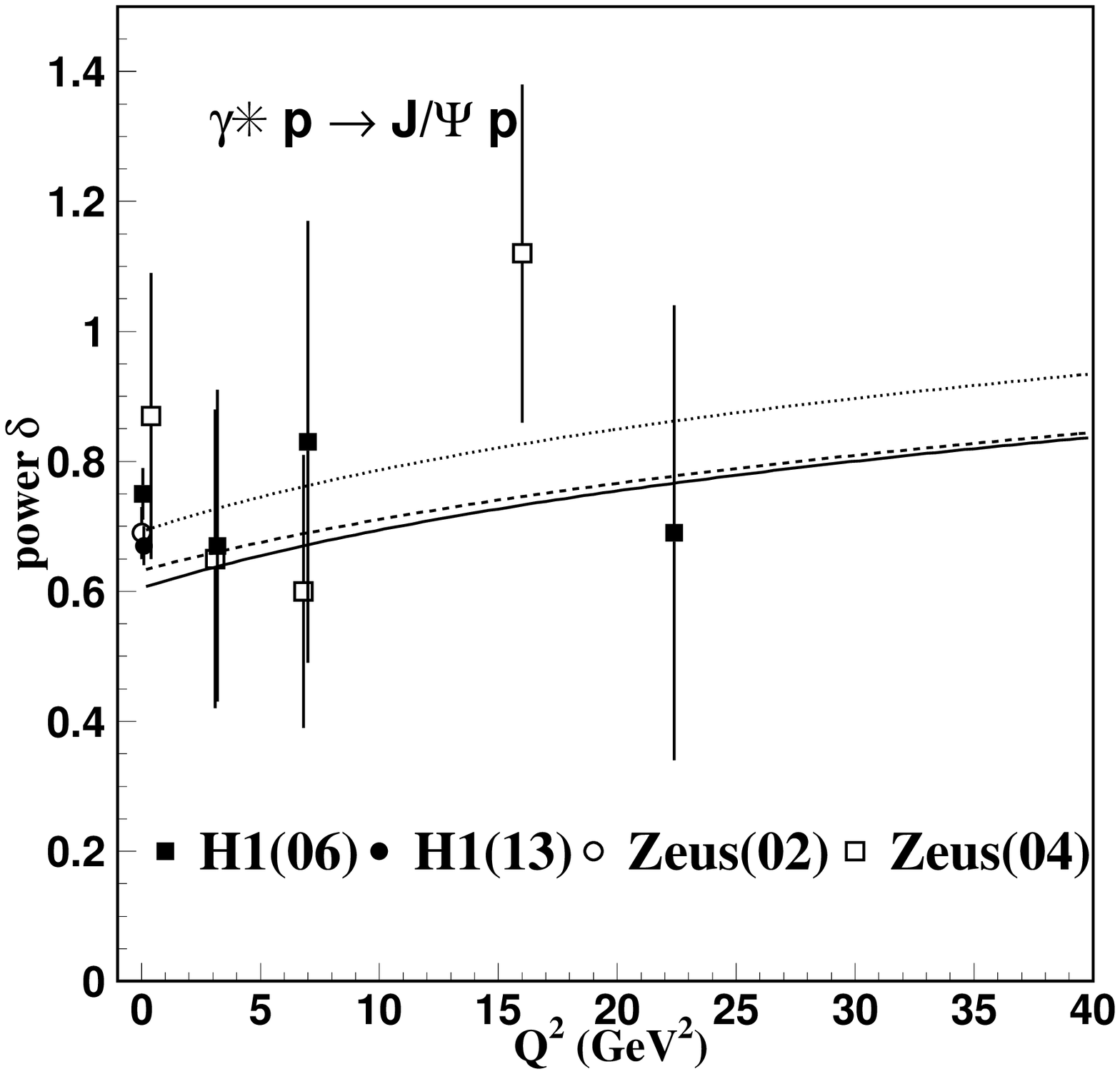}
\includegraphics*[width=8cm]{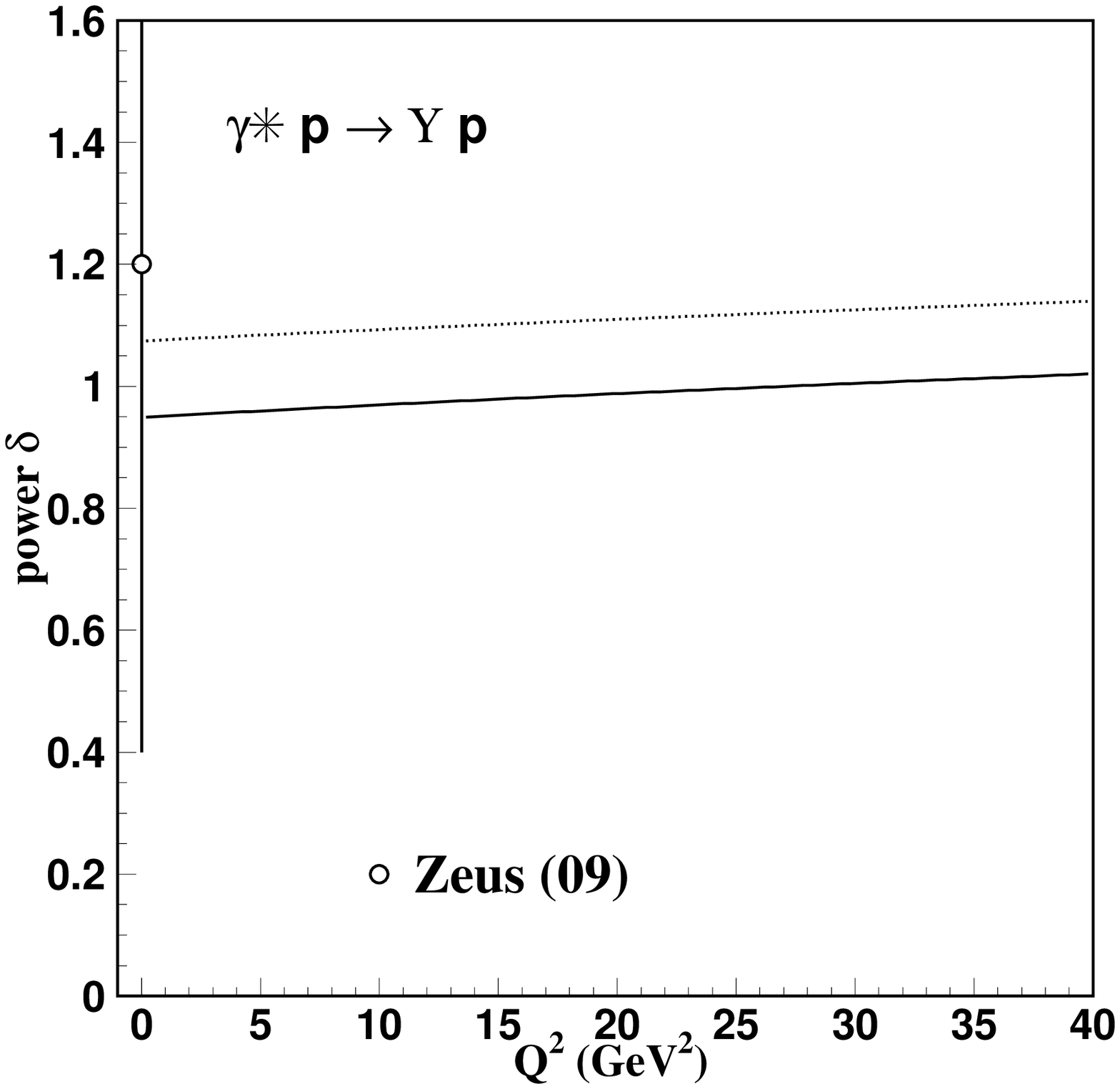}
\end{center}
\caption{ \label{deltas}Predictions for the power $\delta$ and
$\delta_{\rm int} $   as function of $Q^2$ for vector meson
electroproduction  and experimental values for this quantity:  a)
$\rho$-production \cite{Zeus(98),Zeus(99),Zeus(07),H1(00),H1(10)};
 b) $\phi$-production \cite{Zeus(05),H1(10)} ;
c) $J/\psi$-production \cite{H1(06),H1(13),Zeus(02),Zeus(04)}; d)
$\Upsilon$-production \cite{Zeus(09)}.
 The solid  line is
$\delta_{\rm int}$, Eq.\req{delta-int}, including  the shrinkage
correction, and  the dashed line is $\delta$, Eq. \req{meq}, as 
obtained with the quadratic dipole cross section. The power
$\delta$ for the special dipole model \cite{Jeong:2014mla} shown
in Fig.\ref{dipole-cross} is represented by the dotted line,
without shrinkage corrections. }
\end{figure*}

In Fig.\ref{deltas} we show   theoretical predictions and
experimental results for the powers $\delta$ and $\delta_{\rm int}$, that is without and with shrinkage correction,   as a function
of the  photon virtuality $Q^2$  for 
unpolarized elastic
production of all vector mesons in the ground state, the theoretical results for
$\omega$ meson production are not distinguishable from those of
$\rho$ production.  The long-dashed curves represent the
uncorrected power $\delta(Q^2)$, obtained from Eq.\req{meq}, and
the solid line is $\delta_{\rm int}$, Eq.\req{delta-int} that 
includes  shrinkage corrections. We also show with dotted lines
results based on a scale determination with the rather extreme
dipole cross section \cite{Jeong:2014mla} shown in
Fig.\ref{dipole-cross}.  The theoretical predictions are well
compatible with the experiments. The observed  sharp increase of
the power delta with $Q^2$ near $Q^2=0$ indicates that the rapidly
varying shrinkage correction given by Eq.\req{shrink3} is quite realistic.

\begin{figure*}
\begin{center}
\includegraphics*[width=8cm]{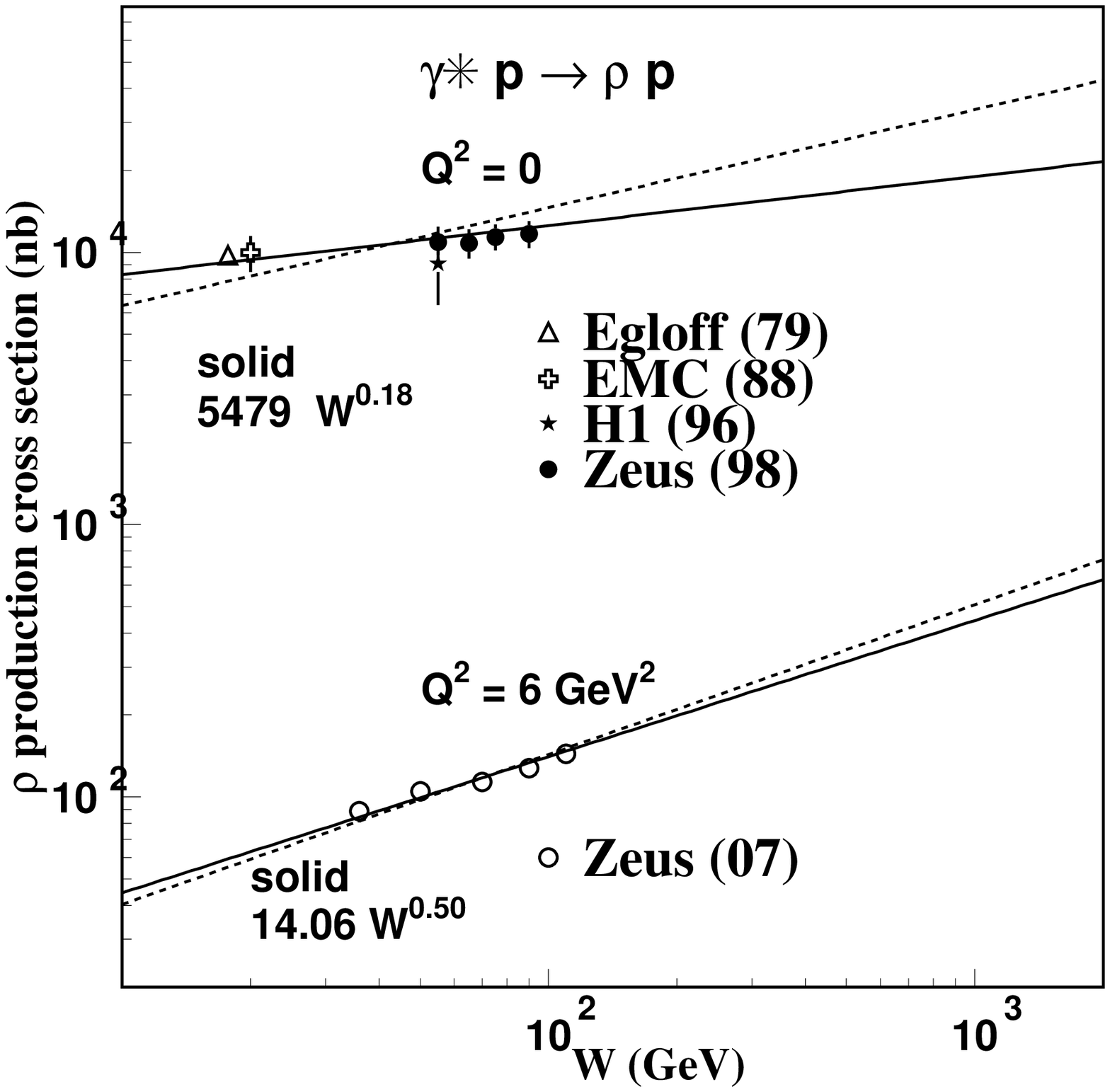}
\includegraphics*[width=8cm]{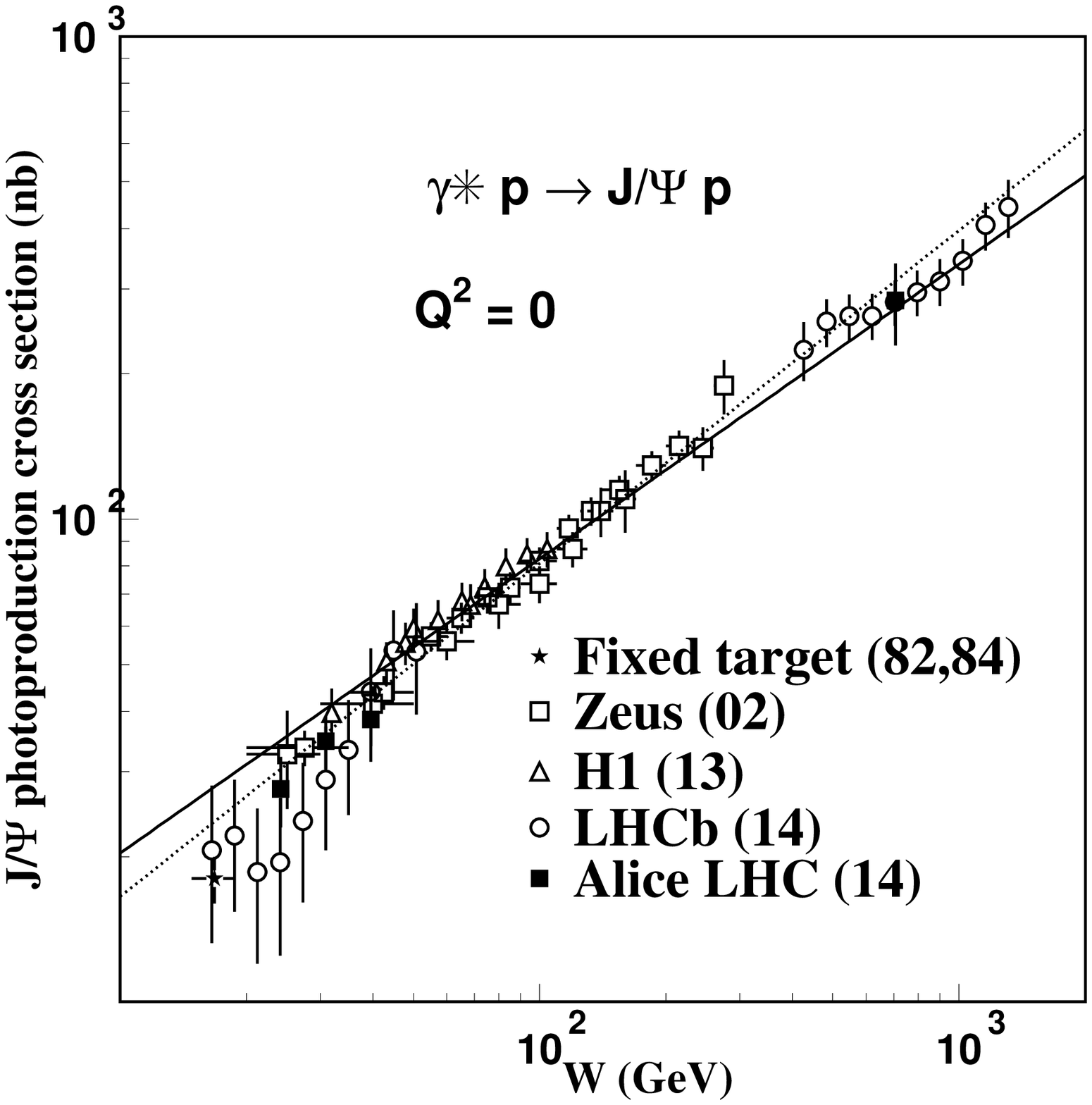}
\end{center}
\caption{ \label{cs}Energy dependence of the
integrated elastic  cross sections  for $\rho$  and $J/\psi$
production.
For $\rho$ production the dashed line shows  the power behaviour without shrinkage 
correction, which is included in Eq.\req{delta-int}, represented by solid lines. 
 For $J/\psi$ production the solid line shows the behaviour with $\delta=0.61$ 
including the (rather small) shrinkage correction, while the dotted line is from 
the extreme dipole cross section of   \cite{Jeong:2014mla} without shrinkage, with 
$\delta=0.69$.
The data for $\rho$ production are from \cite{Egloff,EMC,H1(96),Zeus(98)}
 for $Q^2=0$,  and from 
\cite{Zeus(07)} for $Q^2=6$ . For  $J/\psi$ photoproduction  the data are from
\cite{Zeus(02),Fixed_Target,H1(13),LHCb(14),Alice(14)}. }
\end{figure*}

In in  Fig.\ref{cs}  data and the theoretically predicted energy 
dependence of $\rho$ and $J/\psi$ cross section are displayed. 
According to the model the energy dependence is represented by a single 
power  $c\,W^\eta$ in the full energy range .  The constant $c$ is 
fitted to the data, and the values for the power $\eta$   are given by 
the model, either $\eta=\delta(Q^2)$ or $\eta=\delta_{\rm int}(Q^2)$. For 
$\rho$ production (left hand side)  we show both the results with 
shrinkage correction, Eq.\req{delta-int}, in solid lines, and without shrinkage 
correction, Eq.\req{meq}, in dashed lines. At $Q^2=0$ we have used $\delta(0)=0.36$ 
corresponding to the soft pomeron intercept 1.09. The shrinkage correction 
reduces this vcalue to $\delta_{\rm int}=0.18$.
 The data are clearly compatible with experiment, and it is expected that further 
data at LHC energies will be bring decisive test for the single power behaviour. 

The plot of $J/\psi$ photoproduction on the right hand side   
includes the most recent LHC data.  Here the influence 
of the shrinking correction is small and $\delta(0) = 0.63$, 
$\delta_{\rm shrink}(0) = 0.61$ (solid line).  To exhibit 
the stability of the predictions of the universality conjecture, 
the dotted line gives the prediction from the extreme dipole model 
of \cite{Jeong:2014mla}, displayed in Fig. \ref{dipole-cross}.

  Parameters of the curves $C\,W^{\eta}$  for $J/\psi$  photoproduction 
in some cases are as follows:
 
 theory with shrinkage correction Eq.\req{delta-int}: $\eta= \delta_{\rm int}(0)=0.61,\; C=4.99,\,\chi^2 =1.35$ ; 

 theory without shrinkage correction \req{meq}: $\eta= \delta(0)=0.63,\; C=4.46   ,\,\chi^2 =1.042 $;

theory with dipole model of \cite{Jeong:2014mla} without shrinkage correction: 
$\eta= \delta_{\rm int}(0)=0.69,\; C=3.37   ,\,\chi^2 =0.84$ ;  

free fit with open power: $\eta= 0.68,\, C=3.61, \, \chi^2=0.81$. 

The transverse and longitudinal wave functions are different and
therefore we obtain  different scales for the respective cross sections. 
This leads to different energy behaviour for the two polarizations and the ratio
$R=\si_L/\si_T$ has the power behaviour 
\beq
\label{longtrans} R= \frac{\si_L}{\si_T}= A\, W^{\delta_{\rm R}}~, 
\enq 
with 
\beq \label{lt2} \delta_{ R} = \delta_L- \delta_T ~ .
\enq 
The values of $\delta_L$ and $\delta_T$  are determined by
Eq.\req{meq} with the constants 
$a_{\rm VM,long},\, a_{\rm VM,trans},\,b_{\rm VM,long},\,b_{\rm VM,trans}$ 
of Table \ref{res}. The
experimental errors for the ratio $R$ are quite large and also the theoretical
uncertainties  in the small differences between $\delta_L$ and
$\de_T$ are  large.

\begin{figure*}[h]
\begin{center}
\includegraphics*[width=8cm]{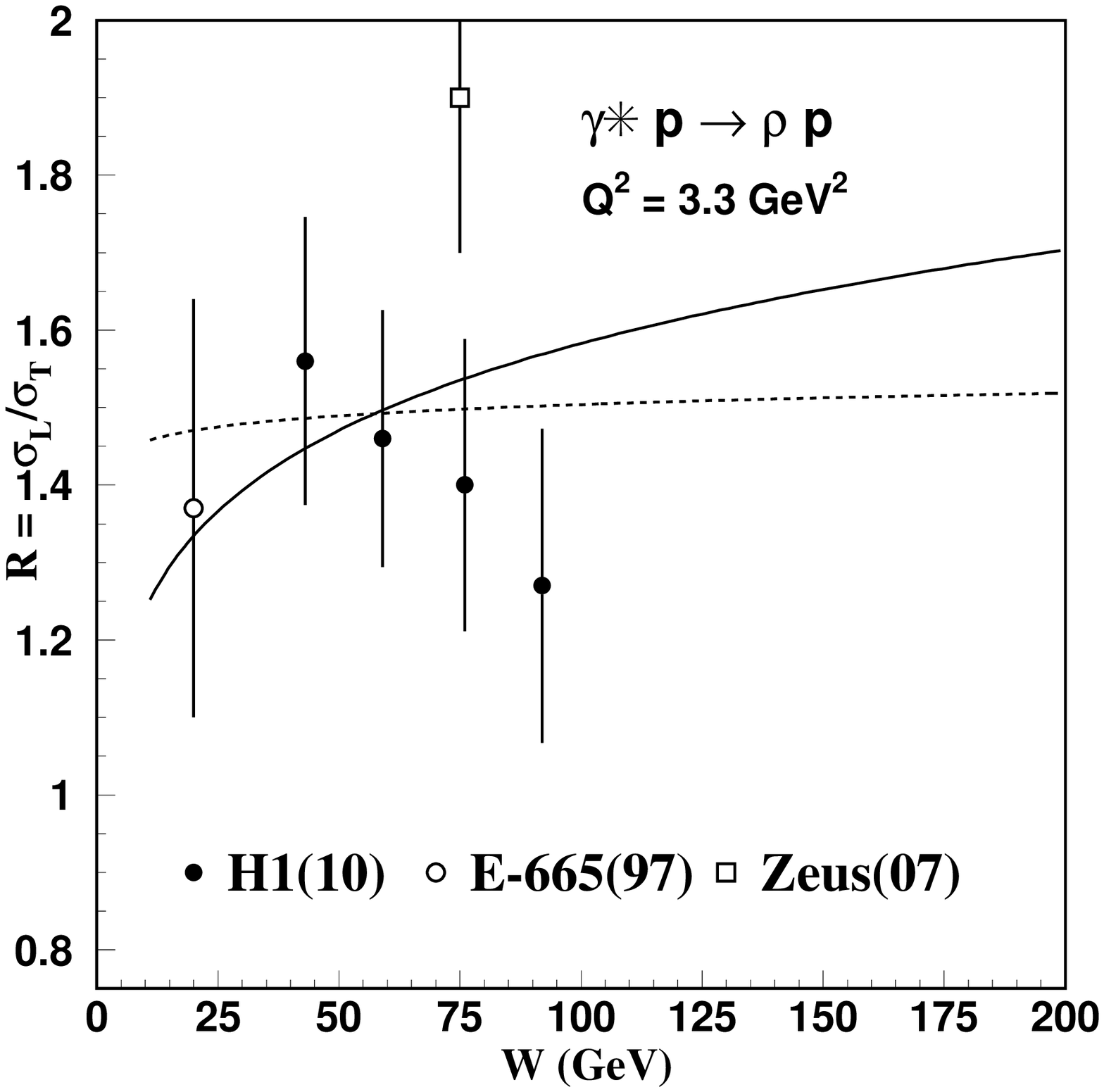}
\includegraphics*[width=8cm]{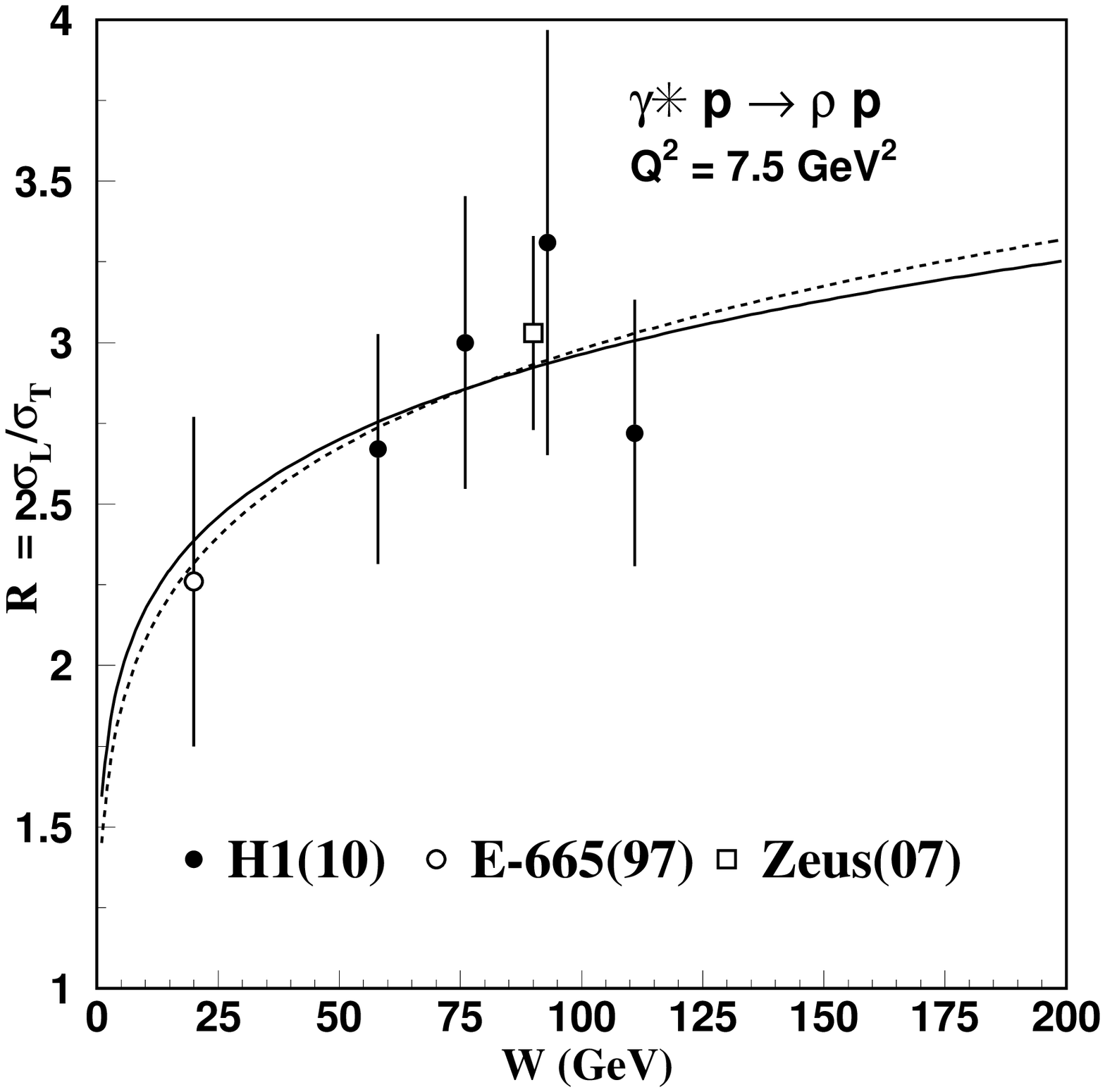}
\includegraphics*[width=8cm]{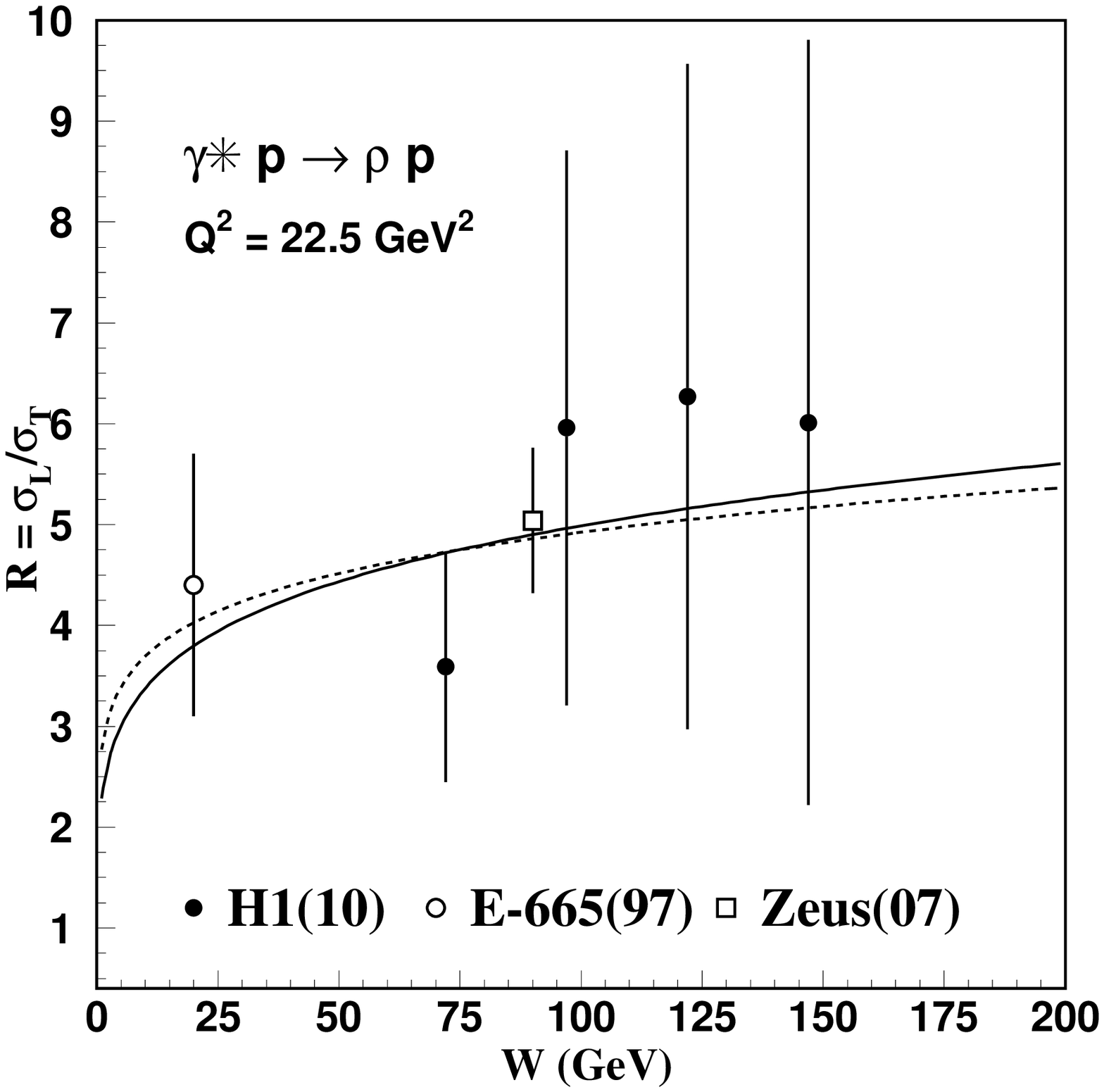}
\includegraphics*[width=8cm]{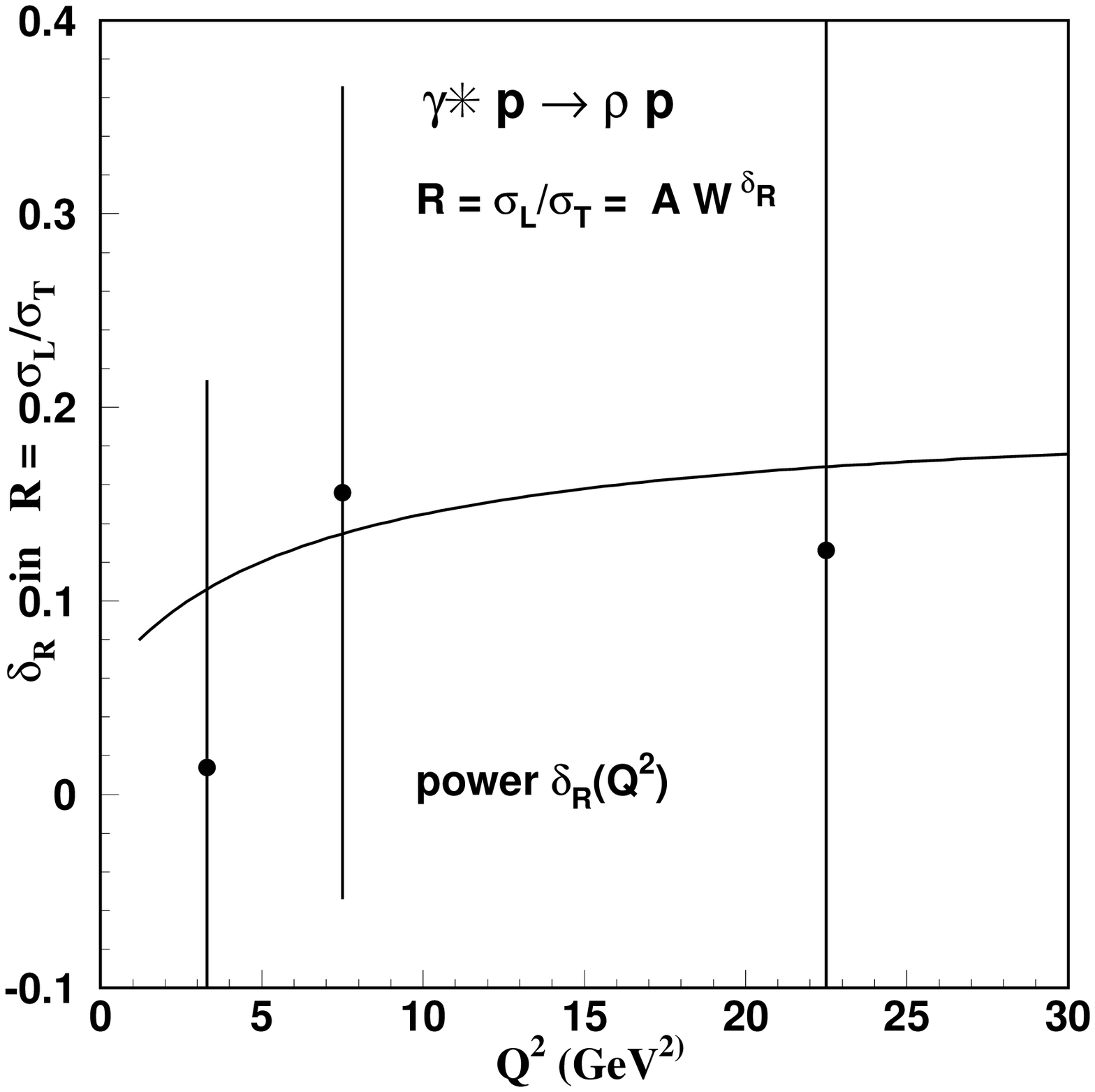}
\end{center}
\caption{ \label{ratios-fig} a)-c): Energy dependence of the
longitudinal to transverse cross section of $\rho$
electroproduction for $Q^2=3.3, 7.5$ and 22.5 GeV$^2$. The solid
lines have the energy dependence predicted according to
Eq.\req{longtrans} with Eq.\req{meq}, and  the dotted lines are free 
best fits
to the data. The data are from \cite{Zeus(07),H1(10),E665(97)} d)
Power $\delta_R$ for the energy dependence of the ratio as
function of $Q^2$. The solid line is  $\delta_L-\delta_T$ from
Eq.\req{meq}, the points with error bars are from the best fits to the
data presented in the plots a)-c).}
\end{figure*}

In Fig. \ref{ratios-fig} a) - c) we show  data
\cite{Zeus(07),H1(10),E665(97)}
 for the energy dependence of the polarization ratios $R=\si_L/\si_T$
for three values of $Q^2$. The solid line are the theoretical
predictions according to Eqs. \ref{meq},\ref{longtrans},\ref{lt2}. The multiplicative
constant $A$ in Eq. \req{longtrans} is fitted freely. We also show in 
dashed lines the results of  free fits to the data with  unconstrained
$A$ and $\delta_R$. At $Q^2=$  7.5 and 22.5 GeV$^2$ the  model
gives good agreement for the energy dependence of the ratio
$R$. In the last plot of the set, 
the data and the theoretical predictions   for the power coefficients as
functions of $Q^2$ are compared directly.

\clearpage

\section{Summary and Conclusions  \label{final}}

We   present  a simple  phenomenological model assuming that
single pomeron exchange determines electromagnetically induced
diffractive processes. The main consequence of this approach is
that the high energy behaviour of each process is determined by a
single power $W^\eta$. Recent results for $J/\psi$
photoproduction from LHC \cite{LHCb(14),Alice(14)} indicate such a
behaviour, which will be tested with more measurements 
of diffractive electromagnetic processes at LHC.

We furthermore assume that the intercept is determined by the
scale of the reaction , which is related to the extension of the
dipole wave functions, as illustrated in Fig.~\ref{dipole}. This 
assumption is inspired by a holographic model which predicts a 
scale dependence of the pomeron slope~\cite{Brower:2006ea}, and is 
here extended to a model where also the intercept  is scale 
dependent. This assumption is necessary if one assumes 
a single pomeron
exchange, since in photon-induced electromagnetic diffraction
processes the energy behaviour depends strongly   on the photon
virtuality. The scale for vector meson production is in our
approach determined  by the extension parameter $\zeta$ of light
cone wave functions~\cite{lep80}  and we determine the scale
dependence of the pomeron intercept by the energy dependence of the
structure function $F_2$ at fixed $Q^2$. We have tested two methods for
the extraction of this scale and also two extremely different
dipole cross sections, illustrated in  Fig. \ref{dipole-cross}. 
It turns out that the uncertainties induced by the different 
methods of definition of the scale
are smaller than present experimental errors.

The assumption of a universal scale, determined essentially by
the effective dipole size of the virtual photon, is well
compatible with the present data. Specifically the energy
dependences of $J/\psi$ and $\Upsilon$ production are successfully
predicted.  A decisive test of the additional hypothesis of an
intercept dependence on a universal scale will also be possible as
data for more reactions in the LHC energy  range become  available.

\begin{figure}[h!]
\begin{center}
\includegraphics[width=8.5cm]{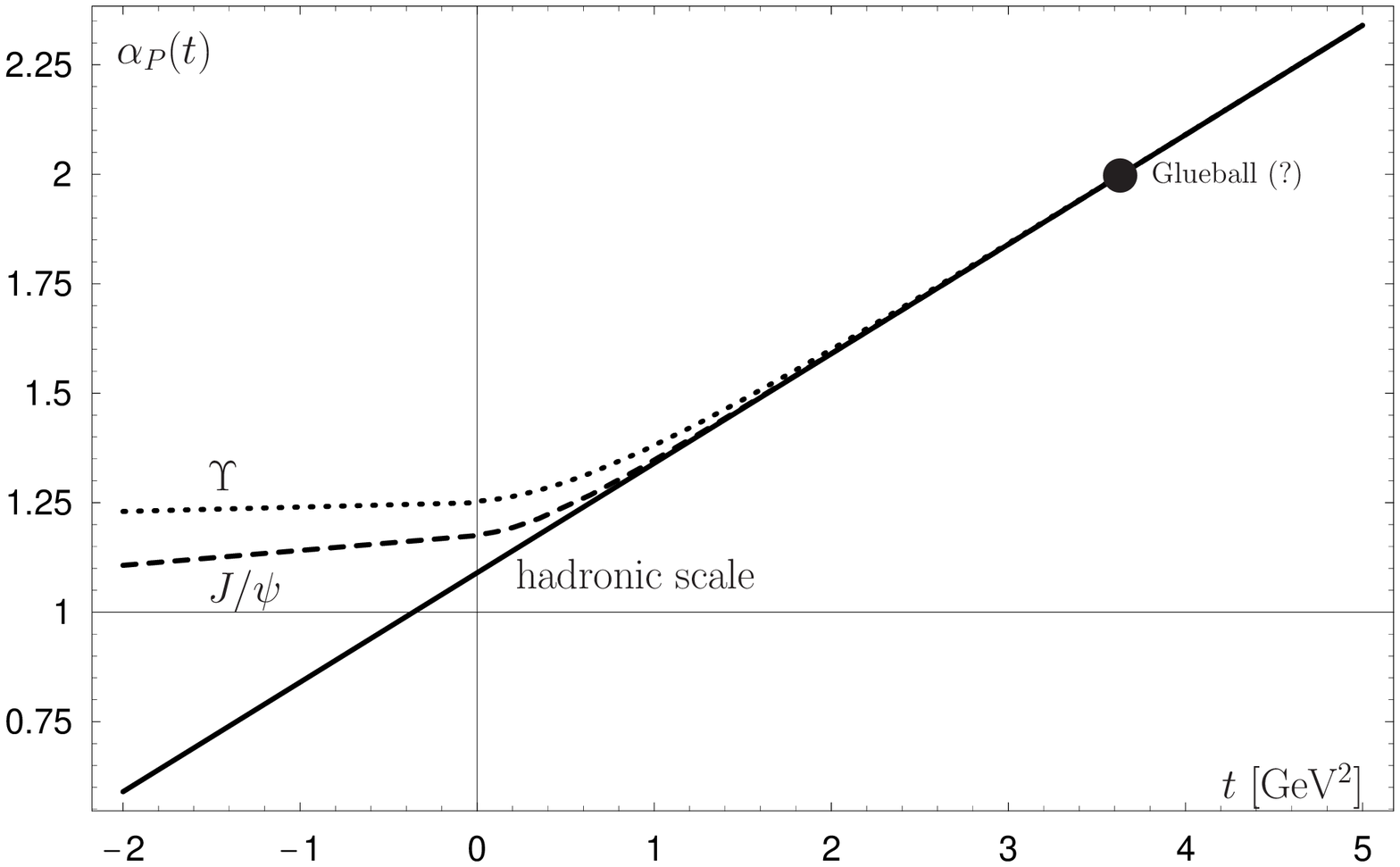}
\end{center}
 \caption{\label{traj} Scenario for scale dependent pomeron trajectories: 
the solid line is the  trajectory relevant at hadronic scales (soft pomeron), 
and the  dashed and dotted  lines represent  the trajectories for $J/\psi$ and  
and  $\Upsilon$ photoproduction, respectively.}
\end{figure}

Also the concept of a scale dependent slope of the pomeron trajectory
~\cite{Brower:2006ea} is well compatible with the data \cite{H1(10)}, 
as shown in Fig.~\ref{cs}. A possible scenario for trajectories   
for $J/\psi$ and $\Upsilon$  photoproduction together with the conventional 
soft pomeron trajectory  is displayed in Fig.~\ref{traj}. The intercept 
$\al_P(0)$ and the slope for $t<0$ are fixed by the model, see Eqs.\req{meq},
\req{shrink3}. For $ t \gg 0$  where glueball states may be on the trajectory, 
the hadronic confinement  scale becomes relevant and there it should coincide 
with the soft pomeron, that is the pomeron trajectory relevant for hadronic 
scattering.

The model is inspired by AdS/CFT type models which yield a scale dependent 
slope of the pomeron trajectory and, in a typical bottom-up approach, it is  
based on the phenomenological input of a scale dependent intercept. A real
challenge would be to show, at least qualitatively, how a scale
dependent intercept can emerge from more fundamental principles.

\begin{acknowledgments}
It is a pleasure to thank  Guy de T\'eramond, Carlo Ewerz, and
Otto Nachtmann for numerous constructive critical suggestions and
remarks. One of the authors, E.F.,  wishes to thank the Brazilian
agencies CNPq, PRONEX, CAPES and FAPERJ for financial support.
\end{acknowledgments}

\clearpage


\begin{thebibliography}{99}

\bibitem{Gribov:1972ri}
  V.~N.~Gribov and L.~N.~Lipatov,
  Sov.\ J.\ Nucl.\ Phys.\  {\bf 15},  438 (1972)
   [Yad.\ Fiz.\  {\bf 15} (1972) 781] 

\bibitem{Altarelli:1977zs}
  G.~Altarelli and G.~Parisi,
  Nucl.\ Phys.\ B {\bf 126}, 298  (1977)  

\bibitem{Dokshitzer:1977sg}
  Y.~L.~Dokshitzer,
  Sov.\ Phys.\ JETP {\bf 46}, 641 (1977)  
   [Zh.\ Eksp.\ Teor.\ Fiz.\  {\bf 73}, 1216 (1977)] 


\bibitem{Donnachie:1992ny}
  A.~Donnachie and P.~V.~Landshoff,
  Phys.\ Lett.\ B {\bf 296}, 227 (1992) 

\bibitem{Nikolaev:1991et}
  N.~Nikolaev and B.~G.~Zakharov,
  Z.\ Phys.\ C {\bf 53}, 331 (1992)  


\bibitem{Ewerz:2006vd}
  C.~Ewerz and O.~Nachtmann,
  Annals Phys.  {\bf 322}  1635, 1670 (2007)
  [hep-ph/0604087].

\bibitem{Ewerz:2007md}
  C.~Ewerz, A.~von Manteuffel and O.~Nachtmann,
  Phys.\ Rev.\ D {\bf 77} (2008) 074022
  [arXiv:0708.3455 [hep-ph]].

\bibitem{Fadin:1975cb}
  V.~S.~Fadin, E.~A.~Kuraev and L.~N.~Lipatov,
  Phys.\ Lett.\ B {\bf 60}, 50 (1975) 


\bibitem{Kuraev:1977fs}
  E.~A.~Kuraev, L.~N.~Lipatov and V.~S.~Fadin,
  Sov.\ Phys.\ JETP {\bf 45}, 199 (1977)  
   [Zh.\ Eksp.\ Teor.\ Fiz.\  {\bf 72}, 377 (1977)] 

\bibitem{Balitsky:1978ic}
  I.~I.~Balitsky and L.~N.~Lipatov,
  Sov.\ J.\ Nucl.\ Phys.\  {\bf 28}  822 (1978)
   [Yad.\ Fiz.\  {\bf 28}, 1597 (1978)] 

\bibitem{Ciafaloni:1998gs}
  M.~Ciafaloni and G.~Camici,
  Phys.\ Lett.\ B {\bf 430}, 349  (1998) 
  [hep-ph/9803389].

\bibitem{Fadin:1998py}
  V.~S.~Fadin and L.~N.~Lipatov,
  Phys.\ Lett.\ B {\bf 429}, 127 (1998) , 
  [hep-ph/9802290].
\bibitem{Donnachie:1998gm}
  A.~Donnachie and P.~V.~Landshoff,
  Phys.\ Lett.\ B {\bf 437}, 408 (1998)

\bibitem{Dosch:1994ym} 
  H.~G.~Dosch, E.~Ferreira and A.~Kramer,
  Phys.\ Rev.\ D {\bf 50}, 1992 (1994)
  [hep-ph/9405237].


\bibitem{Dosch:1997nw}
  H.~G.~Dosch, T.~Gousset and H.~J.~Pirner,
  Phys.\ Rev.\ D {\bf 57} 1666 (1998) 
  [hep-ph/9707264].

\bibitem{Kulzinger:1998hw}
  G.~Kulzinger, H.~G.~Dosch and H.~J.~Pirner,
  Eur.\ Phys.\ J.\ C {\bf 7}, 73 (1999)  
  [hep-ph/9806352]


\bibitem{Donnachie:1999kp}
  A.~Donnachie, H.~G.~Dosch and M.~Rueter,
  Eur.\ Phys.\ J.\ C {\bf 13},  141 (2000)


\bibitem{Donnachie:2000px}
  A.~Donnachie and H.~G.~Dosch,
  Phys.\ Lett.\ B {\bf 502}, 74 (2001) ;  Phys.\ Rev.\ D {\bf 65}, 014019 (2002)



\bibitem{Dosch:2002ig}
  H.~G.~Dosch and E.~Ferreira,
  Eur.\ Phys.\ J.\ C {\bf 29}, 45 (2003)  

\bibitem{Dosch:2006kz}
  H.~G.~Dosch and E.~Ferreira,
  Eur.\ Phys.\ J.\ C {\bf 51}, 83  (2007) 
\bibitem{Baltar:2009vp}
  V.~L.~Baltar, H.~G.~Dosch and E.~Ferreira,
  Int.\ J.\ Mod.\ Phys.\ A {\bf 26}, 2125  (2011)  
 
 \bibitem{LHCb(14)} R. Aaij et al. [LHCb Coll.],  J. Phys. G {\bf 41} ,   055002 (2014) 

\bibitem{Alice(14)} B. Abelev et al. [Alice Coll.], Phys. Rev. Lett. {\bf 113},   232504 (2014) 
 
 \bibitem{Maldacena:1997re}
 J.~M.~Maldacena,
 {Int.\ J.\ Theor.\ Phys.\  {\bf 38}, 1113 (1999)}
  
   \bibitem{Gubser:1998bc}
  S.~S.~Gubser, I.~R.~Klebanov and A.~M.~Polyakov,
{Phys.\ Lett.\ B {\bf 428}, 105 (1998)}

\bibitem{Witten:1998qj}
  E.~Witten,
  ``Anti-de Sitter space and holography,''
  Adv.\ Theor.\ Math.\ Phys.\  {\bf 2}, 253 (1998)
 
\bibitem{Brower:2006ea}
  R.~C.~Brower, J.~Polchinski, M.~J.~Strassler and C.~I.~Tan,
  JHEP {\bf 0712}, 005 (2007)  


\bibitem{Hatta:2007he}
  Y.~Hatta, E.~Iancu and A.~H.~Mueller,
  JHEP {\bf 0801}, 026 (2008) 
  [arXiv:0710.2148 [hep-th]].

\bibitem{Brower:2006xx}
A much more elaborated version of this simple model is given in \cite{Brower:2006ea}, but the simple form seems to be a god starting point for a phenomenological bottom-up approach

\bibitem{Brodsky:2006uqa}
  S.~J.~Brodsky and G.~F.~de Teramond,
  Phys.\ Rev.\ Lett.\  {\bf 96} (2006) 201601
  [hep-ph/0602252].

\bibitem{deTeramond:2008ht}
  G.~F.~de Teramond and S.~J.~Brodsky,
  Phys.\ Rev.\ Lett.\  {\bf 102}, 081601 (2009) 

\bibitem{Brodsky:2014yha}
  S.~J.~Brodsky, G.~F.~de Teramond, H.~G.~Dosch and J.~Erlich,
  arXiv:1407.8131 [hep-ph] ; Phys. Rep. C , to appear.

\bibitem{Forshaw:2013oaa}
  J.~Forshaw and R.~Sandapen,
  PoS DIS {\bf 2013}, 089 (2013)  
  [arXiv:1305.3768 [hep-ph]] 

\bibitem{lep80} G.P. Lepage and S.J. Brodsky, 
Phys. Rev. D {\bf 22}, 2157 (1980)  

\bibitem{Radescu:2013mka}
  V.~Radescu [H1 and ZEUS Coll.],
  Conf. Proceedings, arXiv:1308.0374 [hep-ex] (2013)
  
\bibitem{H1(10)}  F. D. Aaron et al., JHEP 1005,  032 (2010); [arxiv:0910.5831]

\bibitem{Gilman:1970vi}
  F.~J.~Gilman, J.~Pumplin, A.~Schwimmer and L.~Stodolsky,
  Phys.\ Lett.\ B {\bf 31} 387 (1970) .

\bibitem{Jeong:2014mla}
  Y.~S.~Jeong, C.~S.~Kim, M.~V.~Luu and M.~H.~Reno,
  JHEP {\bf 1411} , 025 (2014) 
  [arXiv:1403.2551 [hep-ph]].

\bibitem{Ewerz:2011ph}
  C.~Ewerz, A.~von Manteuffel and O.~Nachtmann,
  JHEP {\bf 1103}, 062 (2011) 
  [arXiv:1101.0288 [hep-ph]].

\bibitem{Agashe:2014kda}
  K.~A.~Olive {\it et al.}  [Particle Data Group Coll.],
  Chin.\ Phys.\ C {\bf 38} 090001   (2014) 

\bibitem{Adloff:2001rw}
  C.~Adloff   et al.   [H1 Coll.],
  Phys.\ Lett.\ B {\bf 520} , 183 (2001) 
\bibitem{Zeus(11)}  S. Chekanov et al. [Zeus Coll.],  
   Phys.\ Lett.\ B {\bf 697}, 184 (2011)  
\bibitem{Zeus(98)} J.Breitweg et al.,  Eur.\ Phys.\ J.\ C {\bf 2}, 247  (1998) 
\bibitem{Zeus(99)} M. Derrick et al., Eur.\ Phys.\ J.\ C {\bf 6}, 603  (1999) 
\bibitem{Zeus(07)}  S. Chekanov , PMC Physics  A   1:6 (2007)   
\bibitem{H1(00)} C. Adloff et al., Eur.\ Phys.\ J.\ C {\bf 13}, 371 (2000)
      
 \bibitem{Zeus(05)}  S. Chekanov et al. Nucl.Phys B {\bf  718 } , 3   (2005)
\bibitem{H1(06)} A. Aktas et al., ,  Eur. Phys. J. C {\bf 46},   585   (2006)  
\bibitem{H1(13)} C. Alexa et al. ,  Eur. Phys. J. C {\bf 73} , 2466 (2013)  
\bibitem{Zeus(02)} S. Chekanov et al. ,  Eur. Phys. J. C {\bf  24}, 345  (2002)   
\bibitem{Zeus(04)}  S. Chekanov et al. ,  Nucl. Phys. B {\bf 695}, 3  (2004)  
\bibitem{Zeus(09)}  S. Chekanov et al. ,  Phys. Lett. B {\bf 680}, 4   (2009)  
  \bibitem{Egloff} R.M. Egloff, P.J.Davis, G.J.Luste, J.F.Martin and J.D.Prentice , 
  Phys. Rev. Lett. {\bf 43} , 657 (1979)   
\bibitem{EMC}   J.J. Aubert et al. [EMC Coll.] ,  Phys. Lett B {\bf 161}, 203  (1985) ; 
  J. Ashman et al. [EMC Coll.] ,  Zeit.Phys. C {\bf  39}, 169 (1988) 
 \bibitem{H1(96)}  S. Aid et al., [H1 Coll.],  Nucl.Phys. B {\bf 463} 3 (1996) 
 \bibitem{Fixed_Target}                                 
 B.H. Denby et al., Phys. Rev. Lett. {\bf 52} , 795  (1984)   ; 
M.E. Binkley et al., Phys. Rev. Lett. {\bf 48}, 73  (1982)  
   
 
\bibitem{H1(13)} C. Alexa et al. [H1 Coll.], Eur. Phys. J. C {\bf 73} , 2466 (2013)  

\bibitem{E665(97)}  M. R. Adams et al.,  Zeit. Phys.  C 74 , 237 (1997)

\end{thebibliography}
\end{document}